\documentclass[onecolumn,authoryear]{els-mrw} 

\usepackage{amsmath,amssymb,amsfonts,amsthm,makeidx,graphicx}
\usepackage{txfonts}
\usepackage{helvet}
\usepackage{wasysym}
\usepackage{comment}

\def\bp{$\beta$\,Pic}
\def\oum{1l/'Oumuamua}
\def\2l{2l/Borisov}
\def\bp{$\beta$\,Pic}
\newcommand{\aap}{Astronomy \& Astrophysics}
\newcommand{\nat}{Nature}
\newcommand{\pasp}{Publications of the ASP}
\newcommand{\aj}{The Astronomical Journal}
\newcommand{\apj}{The Astrophysical Journal}
\newcommand{\apjl}{The Astrophysical Journal Letters}
\newcommand{\araa}{Annual Review of Astronomy and Astrophysics}
\newcommand{\aaps}{Astrophysical \& Astrophysics Supplement Series}
\newcommand{\mnras}{Monthly Notices of the Royal Astronomical Society}

\begin{document}

\chapter{Exocomets, exoasteroids and exomoons}\label{chap1}

\author[1]{Paul A. Str{\o}m}%

\address[1]{\orgname{University of Warwick}, \orgdiv{Department of Physics}, \orgaddress{CV4 7AL, Coventry, UK}}


\maketitle

\begin{abstract}[Abstract]

Comets, asteroids and moons that orbit stars and planets exterior to our solar system are prefixed with "exo". While the existence of these objects is certain, our understanding of their physical properties, composition, and diversity is still in its infancy, especially when compared to similar objects within our own solar system. This chapter introduces the topics of exocomets, exoasteroids, and exomoons, putting in context three emerging subfields in astronomy that, despite being relatively small, have experienced rapid growth over the past decade.
\end{abstract}


\begin{glossary}[Learning Objectives]
By the end of this chapter, you will be familiar with:
\begin{itemize}
    \item What typically defines exocomets, exoasteroids, and exomoons
    \item The evidence supporting their existence
    \item Methods used to observe and characterize these objects
    \item Future prospects for their study
\end{itemize}
\end{glossary}

\begin{glossary}[Glossary]

\term{Argument of periastron:} The angle from a body's ascending node to its periastron, measured in the direction of motion.\\
\term{Ascending node:} Where the orbiting object moves north through the plane of reference.\\
\term{Debris disk:} A type of circumstellar disc made up of dust produced by the fragmentation of larger bodies.\\
\term{Dust:} Small solid particles, typically smaller than 1mm.\\
\term{Exoasteroid:} An asteroid exterior to our solar system.\\
\term{Exocomets:} Comets which orbit other stars than the Sun and which exhibit some form of observable activity such as the release of gas or dust, e.g., through a coma or tails of ions or dust.\\
\term{Exomoon:} A moon which orbits an exoplanet.\\
\term{Exoplanet:} A planet which orbits any star other than our sun.\\
\term{Exozodi:} A short hand term for exozodiacal dust.\\
\term{Exozodiacal dust:} Interplanetary dust orbiting a star other than the Sun.\\
\term{Inclination:} In this context, the angle between the plane of reference and the orbit.\\
\term{Interstellar object:} A substellar object, such as an exocomet, exoplanet or rogue planet which is no longer gravitationally bound to a star.\\
\term{Longitude of ascending node:} The angle from a specified reference direction, called the origin of longitude, to the direction of the ascending node, as measured in a specified reference plane.\\
\term{Periapsis:} The point at which an orbiting object is closest to the center of mass of the body it is orbiting.\\
\term{Radiation pressure:} A force arising from photons hitting a dust grain, which acts in the direction away from the star.\\
\term{Roche radius:} Also referred to the Roche limit or Roche sphere, is the minimum distance at which a celestial body, such as a moon or a satellite, can orbit a larger celestial body without being torn apart by tidal forces.\\
\term{True anomaly:} The angle between the direction of periastron and the current position of the body, as seen from the main focus of the ellipse (the point around which the object orbits).\\
\term{Zodiacal dust:} Interplanetary dust in the solar system.\\

\end{glossary}

\begin{glossary}[Nomenclature]
\begin{tabular}{@{}lp{34pc}@{}}
ALMA & Atacama Large Millimeter/submillimeter Array: A radio telescope array in Chile that observes the universe in millimeter and submillimeter wavelengths.\\
CaII & Singly ionized calcium: A spectral line frequently used to study exocomets.\\
COS & Cosmic Origins Spectrograph: A far-UV spectrograph onboard the Hubble Space Telescope.\\
HEK & Hunt for Exomoons with Kepler\\
HST & Hubble Space Telescope: A space-observatory equipped with several instruments covering a broad wavelength range.\\
HTC & Halley-type comets: Comets with orbital periods of between 20 and 200 years and inclinations extending from zero to more than 90 degrees.\\
PLATO & PLAnetary Transits and Oscillations of stars: An ESA space-based observatory to be launched at the end of 2026.\\
STIS & Space Telescope Imaging Spectrograph: A spectrograph, also with a camera mode, installed on the Hubble Space Telescope.\\
TESS & Transiting Exoplanet Survey Satellite: A space telescope aimed at detecting exoplanets.\\
TTV & Transit Timing Variation: A method for detecting exoplanets by observing variations in the timing of a transit.\\
WD & White Dwarf: A remnant of a dead low mass star.\\
\end{tabular}
\end{glossary}

\newpage

\section{Exocomets}\label{exocomets}
\subsection{Introduction}
Exocomets are comets which exist outside our own solar system and may either orbit other stars or travel between them (interstellar objects). Together with exoasteroids they are regarded as the unused building blocks of planetary systems, having formed during the earliest stages of stellar system evolution. Astronomers are interested in studying them as their composition provides pristine samples of the formation and evolution conditions of early stellar systems. The presence of exocomets can also indicate the possible presence of planets, which may be altering the trajectory of exocomets, resulting in orbits which cause some exocomets to pass by in close proximity to a star. Once sufficiently close, the exocomet nucleus starts to sublimate, resulting in the release of dust and gas, which forms exocometary tails. It is this release of dust and gas which make the exocomets detectable. The exocomets' nuclei, which are typically only a few km in size, are much too small to be directly detected through direct solid body occultation. The coma, which surrounds the nucleus and the tails, can be very large and able to cover a significant fraction of the stellar surface.

Spectroscopic signatures are detected when the starlight is absorbed by the gas released by the exocomets (or created through dust desorption) creating additional variable absorption signatures in the stellar spectrum. These signatures can provide information about the exocomet composition and dynamics (such as the radial velocity and acceleration). Photometric transits are seen when the dust released by the exocomet obscures the stellar light (blocking or reflecting it as opposed to only absorbing it) causing a decrease in the observed brightness of the star. These measurements give insight into the dust properties, length and size of the tail as well as an estimate of the size of the exocomet nucleus.


\subsection{Exocomet composition}\label{sec:exocomet_composition}
The composition of exocomets can be probed in a few ways: through transit observations (when the exocomet passes in front of the star and thus absorbs or blocks out some of the light), by studying the gas in circumstellar discs (through mm-wavelength observations), or by analysing WD spectra which, in some cases, show the signatures of exocometary material having been accreted onto them. Each method is sensitive to detecting different compositional information and are thus complementary. As an example, when conducting transit observations we are probing the coma (and the tail) and getting information about ionised gas surrounding (or trailing behind) the comet. When we detect exocometary gas within a circumstellar disc, such as a debris disc, we are sensitive to detecting molecules which have either been left behind by exocomets or released as they disintegrate or collide. We are also not in this case probing the composition of a single exocomet, but an average of a great number of comets within the disc. Analysis of WD spectra opens up the possibility to study the interior composition as the exocomet is broken apart as it accretes onto the WD.

\subsubsection{Spectroscopically observed transits}
\begin{figure}[t]
\centering
\includegraphics[width=0.7\textwidth]{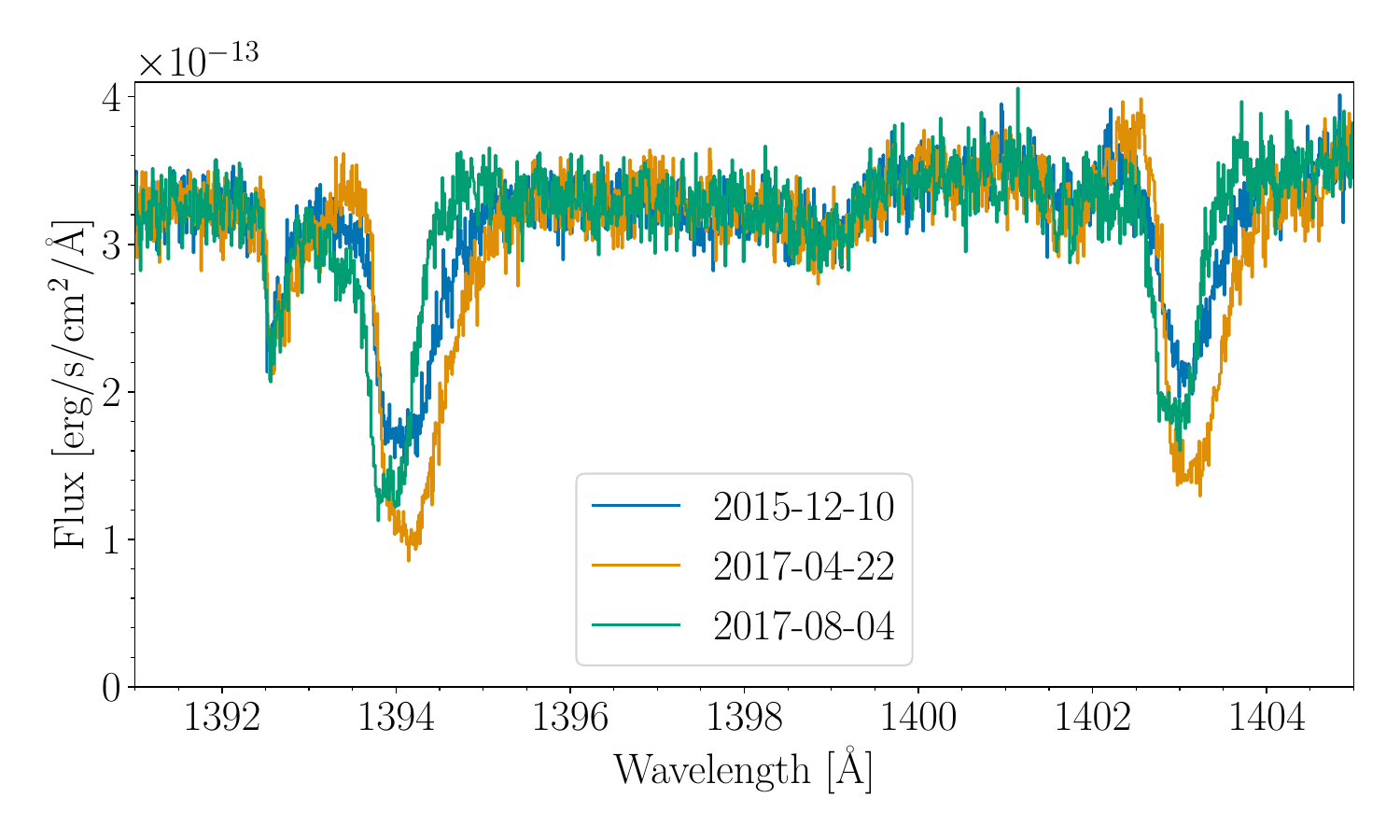}
    \caption{Variable SiIV absorption lines caused by exocomets transiting the star \bp, as observed by \textit{HST}/COS. The variable absorption are seen in both the lines of the SiIV doublet.}
    \label{fig:SiIV}
\end{figure}

The first exocomets were discovered through spectroscopic observations of the Ca II lines of the star \bp\, \citep{ferlet_1987}. Observations of this bright star showed that in addition to the stellar Ca II lines, there were also variable Doppler shifted Ca II absorption lines. The additional absorption was attributed to exocomets whose dust sublimates into a gas as they approached the star and which soon after becomes ionised by UV radiation to produce Ca$^+$ ions. Numerical simulations, which assume a given isotropic dust distribution, gas production rate and size distribution of dust grains are able to reproduce the observational data, supporting the idea that exocomets cause the absorption \citep{beust_1990}. Subsequent observations of the Ca II absorption lines have shown this variability to be omnipresent with variations seen from night to night and even down to timescales of minutes. Fig.\,\ref{fig:SiIV} shows what this additional absorption typically looks like. Since the late 80's thousands of exocomets have been observed, mostly around \bp\, \citep{kiefer_2014}. There have also been numerous exocomet species detected, mostly ions (CaII, FeII, MgII, CII, AlIII and SiIV to mention some); see \citealt{strom_2020} for an overview and references. Although exocomet detection have been claimed for a tens of stars, there is only a small handful of stars which display clear spectroscopic exocomet signatures, 49 Cet and HD 172555 being amongst the few that do so.

Despite being able to detect multiple species, we are unable to measure the quantity of a given element given our fixed vantage point. This is because as an observer we can only measure an absorption signature when the gas lies between the star and us. A consequence of this is that we are unable to detect gas which may have originated from the comet but does not pass our line of sight. What we can measure is the column density which is the density along our line of sight. Although it does not bring us closer to measuring the quantities of exocometary material around the exocomets, is does provide us valuable insights into the composition of exocomets when compared to column density measurements of other species. That is, the ratio of column densities is directly representative of the abundance ratios. We can't say how much iron is in the vicinity of the exocomet, but we can say how much more there is of carbon relative to iron. This is made possible under the assumption that the ions are well mixed in the coma and tail.

Measuring the column densities and abundance ratios of exocomets precisely has proven to be a challenge. Saturated lines, which is one of the biggest challenges, make it hard to get an accurate column density measurement. Another challenge is that the exocometary absorption profile is often influenced by absorption lines from other species at similar wavelengths, making it difficult to determine each species' contribution to the absorption. The problem can be mitigated by modelling multiple absorption lines for the various different species. However, this introduces additional uncertainties, especially for lower resolution spectra where the lines are more likely to be blended. Higher resolution spectra lessens this problem. A limited spectral coverage may also make it harder to determine the correct abundance ratios of species, considering that the column density of a given ion (e.g. Fe II) does not inform us about the column density of other ionisation states of the same element (e.g. Fe I, Fe III, Fe IV). More often than not, only the absorption profiles of one or two ionisation states of any given species will be accessible. Without knowing what ionised state the majority of the elements are in may lead to an incorrect estimate of the total abundance for a particular element (e.g. Fe$_\mathrm{total}$ = Fe + Fe\,II + Fe\,III + Fe\,IV + $\dots$). Elements which are expected to feature prominently, such as H and O, do remain difficult to measure. One reason is that a lot of their lines are only accessible in the far-UV, where the flux from the star is typically very low. Species also present in Earth's upper atmosphere also pose a challenge due to fluorescence, which often produce an emission line very close to or overlapping the measured absorption line.

A new technique has been created, aimed at tackling this problem. Termed the "exocomet curve of growth" technique \citep{vrignaud_2024}, it allows the physical properties of a given species (covering factor, column density) associated with transiting comets to be derived through the study of a large number of absorption lines with different oscillator strengths. This estimate is made possible because the absorption depth increases predictably with the theoretical line strength. This relationship has similar characteristics to \textit{the curve of growth} relationship where an increase in the column density of a material, like interstellar gas, leads to an increase in the equivalent width of the absorption line. The strength of this approach is that is does not require to fit the entire absorption profile of an exocomet; instead, the derivation of the global properties of a given ion (such as Fe II) relies on the measurement of the comet's absorption depth in a large number of lines of this ion with various strengths. With this technique, it is possible to estimate the total covering factor of the comet (i.e, the fraction of the stellar disk it occults), by studying where the absorption depth saturates (the point where no significant increase in absorption depth is seen, despite an increase in line strength). It becomes then possible to estimate the column density of the studied ion in the transiting comet through the study of less saturated lines. The technique has so far only successfully applied to \bp. This is not due to limitation of the technique, but rather a lack of targets as favourable as \bp\, when it comes to exocometary activity coupled with favourable observing conditions (very bright system with a high rate of transiting exocomets).

\subsubsection{Photometrically observed transits}
Dust released by transiting exocomets blocks out light from the host star, resulting in an observed decrease in stellar brightness. The resulting light curve of the exocomet transit often takes on the shape of a saw tooth, yet in other cases may mimic very well the symmetric light curve caused by planetary occultations. There is currently no way to distinguish symmetric exocometary transits from exoplanetary transits, except that exoplanet transits will be periodic, while exocomets will likely not have short enough periods to be re-observed. The exact shape depends on the properties of the star (the radiation pressure and stellar wind) and the exocomet itself (such as the size and orbital characteristics which affect the viewing angle). It is also worth noting, that the dust tail, like the comets of our own solar system, typically has a more curved shape, unlike the ion tail which points away from the star more radially. As the star pushes dust particles away, their orbital period increases causing a lag which creates the curved shape.

The resulting light curve can be modelled both numerically and analytically. Numerical simulations of photometric exocomet occultations was first developed by \cite{lecavelier_1999}. In this approach the photometric variation in the stellar light is calculated for a given dust size distribution around an exocomet by taking into account the optical properties of the dust grains. This includes modelling the spatial distribution of grains and how they are subsequently accelerated away from the star due to radiation pressure. This nonanalytical approach, like any photometric modelling attempt, is highly dependent on various assumptions such as the orbital properties of the exocomets and size and spatial distribution of the grains. When compared to observed light curves, a numerical model excels in providing estimates of physical quantities such as dust production rates, grain size distributions and the distribution of dust in the cometary tail. Analytic approaches have been developed which allow observed transit signatures to be fit. The fit allows one to gain basic information about the exocomet such as an estimate of the dust production rate and nucleus size. There have been numerous approaches aimed at modelling a photometric exocomet signature. Examples include \cite{zieba_2019}, who adapted 1-D model based on the work by \cite{brogi_2012}, which was initially created to model a disintegrating transiting exoplanet, which like an exocomet, exhibits a tail of dust. \cite{kennedy_2019}, used a modified Gaussian which has an exponential tail instead of a Gaussian beyond the transit centre. The model used by \cite{lecavelier_2022} consists of two exponential functions that measured the absorption depth, and from it, infered the exocomet size distribution for the \bp\, system. This particular model can be presented mathematically as:

\begin{equation}
    \frac{\Delta F}{F}(t) = K \left( e^{-\Delta} - e^{\Delta'}\right)
\label{eq_exocomet}
\end{equation}

where 
\[
\Delta = \begin{cases}
\beta(t - t_0) & \text{if } t \geq t_0 \\
0 & \text{if } t \leq t_0
\end{cases}
\]

and 
\[
\Delta' = \begin{cases}
\beta(t - t_0 - \Delta t) & \text{if } t \geq t_0 + \Delta t \\
0 & \text{if } t \leq t_0 + \Delta t
\end{cases}
\]

\begin{figure*}
\includegraphics[width=0.33\columnwidth]{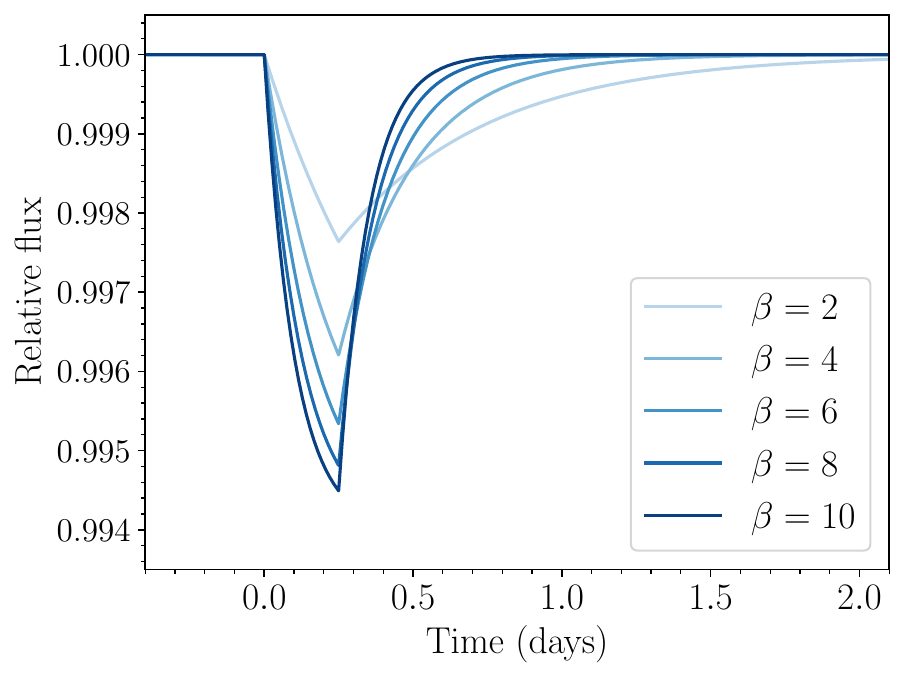}
\includegraphics[width=0.33\columnwidth]{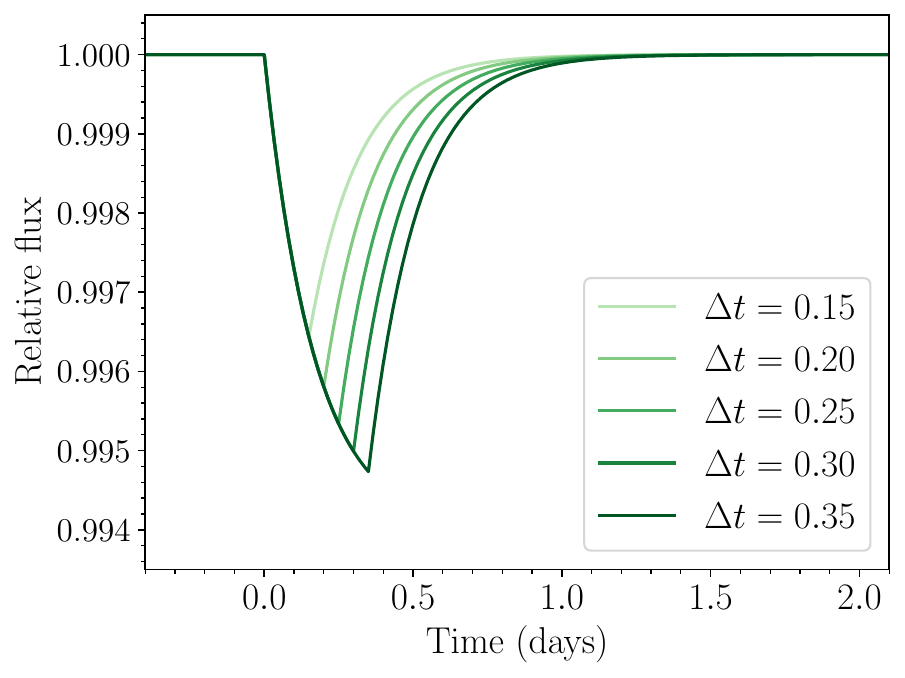}
\includegraphics[width=0.33\columnwidth]{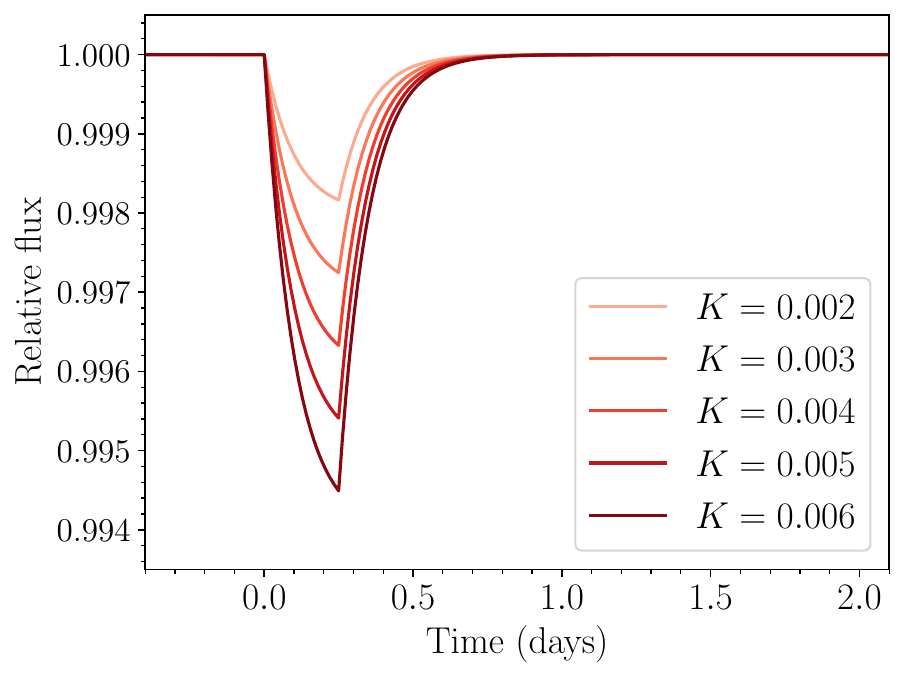}
    \caption{Various exocomet models showing the effects of changing the model parameters $\beta$ (left), $\Delta t$ (middle) and $K$ (right). Changing $\beta$ affects the depth, duration and shape $\Delta t$ affects the depth, and the time of maximum absorption whilst maintaining the overall shape. Changing K only affects the amplitude.}
    \label{fig:model_params}
\end{figure*}

For this model, $K$, is the cloud optical thickness at the leading head of the exocomet. $\Delta t$ is the transit duration which is measured as the time required to cover the chord length of the stellar disc at the transit velocity. $\beta$ is the inverse of the scale length\footnote{Scale length is the characteristic distance over which a physical quantity, like density or brightness, decreases exponentially, typically following the form $f(x) = f_0 e^{-x/L}$, where $L$ represents the distance at which the quantity drops by a factor of $e$.} of the exocomet tail divided by the transit velocity. The start of the transit is given by $t_0$. The effects of varying each of these parameters are shown in Fig\,\ref{fig:model_params}. It was this model which was used to detect and characterise photometric transits seen in the \bp\, observations done with NASA's Transiting Exoplanet Survey Satellite (TESS). Through measuring the absorption depth $AD = K(1-e^{-\beta\Delta t})$ \citealt{lecavelier_2022} inferred the size distribution of exocomets detected in this system and their similarity with the size distribution of comets in our solar system.

As with the spectroscopic observations, \bp\, remains the most well studied target for photometric exocomet transits. There are only a small handful of other stars which exhibit exocometary photometric activity. One of them is HD172555.

The exocometary transits are largely achromatic. However, if the dust particles are sufficiently small then blue light may exhibit an excess extinction of a few percent although this remains to be observed using multi-colour photometry.


\subsubsection{Volatile rich debris discs and exozodiacal dust}

Debris discs are a type of circumstellar discs which are typically rich in dust when compared to gas. They are the exo-analogues to the asteroid and Kuiper belt found in our solar system. It is thought that the dust in these discs come from ongoing collisions and grinding down of asteroid and cometary bodies, collectively called planetesimals. Although these discs are mostly composed of dust, some also contain a significant amount of gas. Over the past decade, the Atacama Large Millimeter/submillimeter Array (ALMA) has detected gas amongst the brighter debris disks, however, most have no detected gas. It remains unknown why some debris disks have detectable gas at all given that this observed gas should have a very short lifetime due to photodisassociation. There are two main competing hypothesis which provide an explanation for why the gas persists in the disk. The first is that the gas is primordial, being left over from the formation from the protoplanetary disc stage, having in some way been shielded from photodisassociation. The second hypothesis is that the gas is 'second generation' gas. That means that the gas has been created at a later stage, such as when planetesimals (including exocomets) collide and grind down within their natal belts releasing volatiles, which quickly becomes gas. The majority of these gaseous debris discs show gas located roughly 100 au from the star, with the age of the systems typically ranging from 10 to a few hundred million years old.

Observations of the CO molecule in particular provides the most compelling support for the planetesimal sublimation hypothesis. As the CO is released from solids via collisions, it is subsequently quickly photodissociated by stellar and interstellar UV, producing C and O, with the process happening on timescales of the order of $100$ years. Thus unless the CO molecule is in some way shielded, it has to be continuously produced for us to be able to observe it. Exocomets, similarly to comets in the solar system, may be rich in CO and thus keep replenishing the CO in debris disks. This may be the case for discs with low CO masses where the rate of CO replenishment agrees with the exocomet destruction rates estimated from continuum observations \citep{cataldi_2023}. Studying the secondary gas can give us an idea of the composition of the solids from which the gas formed. For instance, measurements show that CO ice mass fractions in exocometary belts are comparable to solar system comets, indicating the feasibility of detecting exocometary gas and suggesting similarities in cometary compositions among planetary systems \citep{matra_2017}.

Mid-infrared interferometric observations have shown the presence of warm dust grains present in the habitable regions of exoplanet systems. Referred to as exozodiacal dust or exozodis for short, they resemble the asteroid belt or zodiacal dust in our solar system. In our solar system, comets are the dominant source of the zodiacal dust. As such it seems both possible and likely that the warm exozodiacal dust originates from exocomets. However, around a fifth of main sequence stars show the presence of hot dust close to the star (termed hot exozodis) as detected by near-infrared interferometry observations. Hot dust is unlikely to have formed in situ from a belt close to the star given that such a belt would grind away rapidly. It is more likely the dust comes from further out in the system then travels inwards towards the inner most regions of the star, replenishing the hot dust, which would otherwise get blown away by radiation pressure or end up sublimating into gas. This poses a challenge to explain, since systems which show hot exozodiacal dust don't necessarily show any warm dust and vice versa. The hot dust signatures could persist if the deposition rate is very high or if there is a dust-trapping mechanism, such as gas or magnetic trapping \citep{pearce_2022}. Currently, no model fully explains hot exozodis or their prevalence across various star types and ages, making the presence of this hot dust a mystery.

\subsection{Exocomet orbits}\label{sec:exocomet_orbits}
\begin{figure}[t]
\centering
\includegraphics[width=\textwidth]{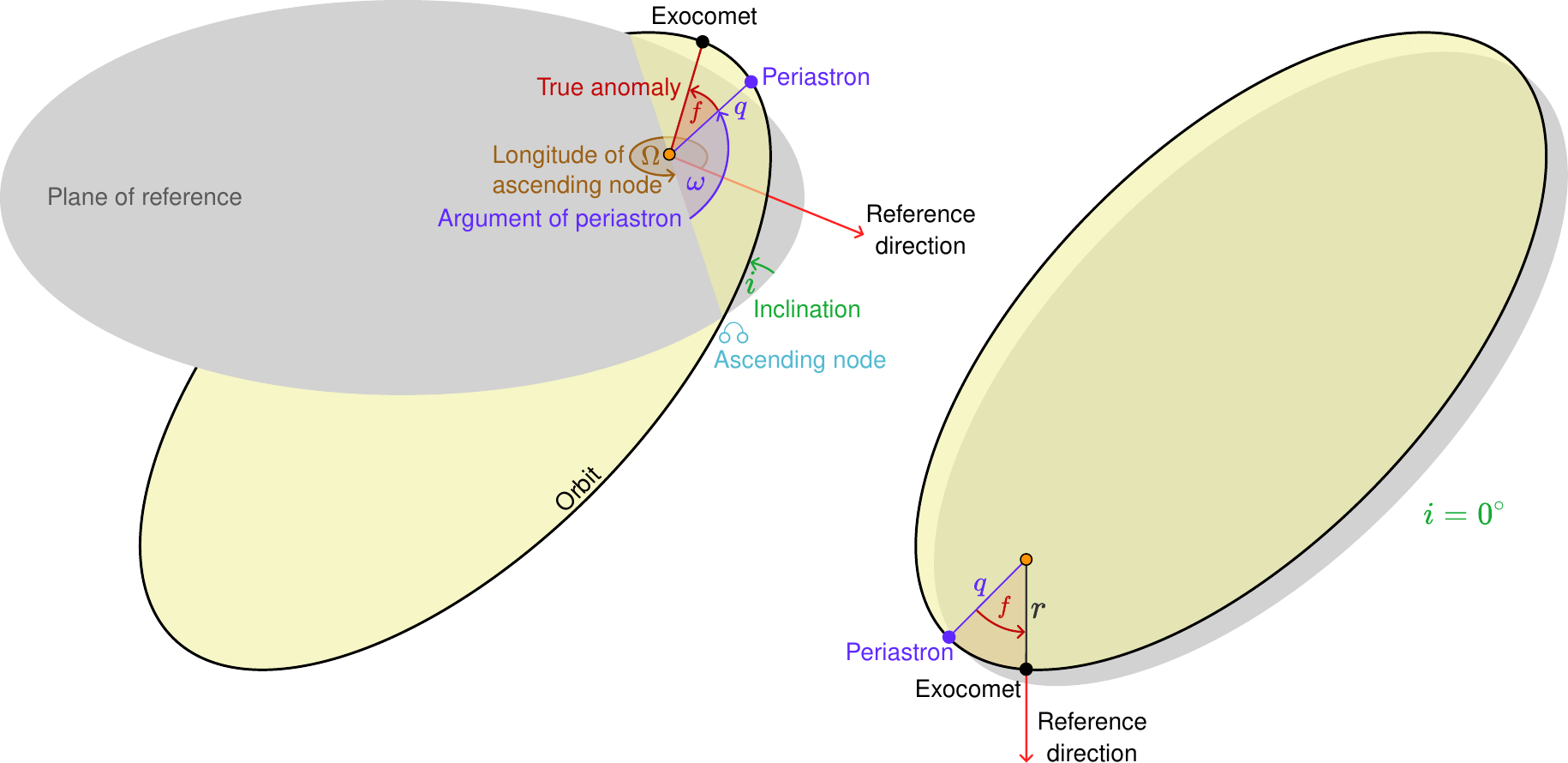}
    \caption{Diagram illustrating and explaining various terms in relation to orbits of celestial bodies. The figure on the right is the case where $i=0^\circ$ making the plane of reference paralell with the orbital plane. It is also rotated such that the reference direction is towards the observer.}
    \label{fig:orbit}
\end{figure}

As for exoplanets, to specify the orbit of an exocomet, five orbital elements, along with the reference time at which they are defined, must be known: the eccentricity $e$, the periastron (closest distance to the star) $q$, inclination $i$, longitude of the ascending node $\Omega$, argument of periastron $\omega$ and a reference time epoch $T_0$. The reference time is typically defined as the time since the object last passed through periastron. Other orbital elements are sometimes given instead, such as the Keplerian elements $a$ (semi-major axis) and the true anomaly $f$. However, as long as the orbit is defined, it is possible to convert between the orbital elements. For instance, one may obtain $q$ if the semi-major axis $a$ and $e$ are known. The six orbital elements describe the shape of the orbit ($e$ and $q$) and the orientation of the orbit in its plane ($i$, $\Omega$ and $\omega$) at a given time $T_0$, which determines the exocomet's position in the orbit. Fig.\,\ref{fig:orbit} illustrates these orbital terms.

The orbital equations can be simplified for transiting exocomets due to the simplified geometric configuration. Since the orbital plane of a transiting exocomet is observed edge-on, we can choose the reference plane to coincide with the orbital plane, reducing the inclination angle to zero ($i=0^\circ$), as illustrated in Fig.\,\ref{fig:orbit}. For a transiting exocomet, it is logical to define the reference direction as pointing toward the observer. With $i=0^\circ$, the ascending node is undefined, so we set $\Omega=0^\circ$. This leaves us with only four elements left to be constrained: $e$, $q$ and $\omega$ for a given reference time $T_0$. Since the ascending node is undefined, $i=0^\circ$, we introduce a new angular term, $\varpi$, known as the longitude of periastron, which measures the angle between the reference direction and the periastron. In our case, where $i=0^\circ$, we have: $\varpi = \Omega + \omega$ and so it follows that $\varpi = \omega$. The angle remains constant over time (assuming no additional perturbations) and so is used to characterise the orbit.

These parameters, which define the orbit, remain constant over time. However, to specify the exocomet's position within the orbit, which changes over time, the true anomaly, $f$, is commonly used instead of $T_0$. The true anomaly is the angle measured from the periastron to the current position of the exocomet, describing its location in the orbit as an angular displacement rather than as a time since periastron passage. At the time of transit, and only then, we have that $f=-\varpi=-\omega$ and so we are left with three remaining elements which govern the shape ($e$), size ($q$) and the orientation of the orbit ($f$). These remaining three parameters show a strong degeneracy with each other, meaning that a single observation can be equally consistent with different combinations of $e$, $q$ and $f$. We have no way of measuring $e$ and thus have to rely on assumptions. An often-used assumption is to set $e=1$, which means the orbits are parabolic. This is reasonable because for large eccentricities close to one, especially near the periastron where the exocomet is observed, there is minimal difference between a parabolic orbit and an orbit with $e\neq1$ but close to 1.

Spectroscopic observations of exocomets give us the most information about exocomet orbits. As an exocomet transits its host star, as seen from our perspective, star light is absorbed by the gas in the coma and tail at certain wavelengths (the wavelength depends on what spectral lines absorb the light). The absorption is seen Doppler shifted with respect to the stellar radial velocity. This radial velocity difference provides information on how fast the exocomet is moving radially relative to the star. In some cases, when the spectral resolution is high enough and the duration of the observations long enough, we may also be able to measure how this Doppler shifted absorption profile changes with time and from that directly measure the radial acceleration/deceleration of the exocomet allowing us to compute the distance between the exocomet and the star as:

\begin{equation}
r = \sqrt{\frac{GM_\star}{dv/dt}}
\label{eq:d}
\end{equation}

\noindent from Newton's law of gravity, where $G$ is the gravitational constant, $M_\star$ the stellar mass and $dv/dt$ the acceleration. Determining $r$ is important because it provides the necessary information to test coma models, which predict how the coma size varies and what species should be present at different distances from the star. As mentioned above our goal is to constrain $e$, $q$ and $f$ or $\varpi$. With the distance calculated and $e$ fixed at $e=1$, we can derive an expression which relates the two remaining parameters: periastron distance $q$ and the true anomaly $f$. We begin with the parabolic conic section, which in polar coordinates is given by:

\begin{equation}
r = \frac{p}{1+e\cos(f)}
\label{eq:parabolic}
\end{equation}

\noindent where p is the semi-latus rectum. For a parabola this is $p=2q$, and with $e=1$ we get:

\begin{equation}
r = \frac{2q}{1+cos(f)}
\label{eq:parabolic}
\end{equation}

\noindent a single equation with two unknowns, $q$ and $f$. To resolve the degeneracy between these two unknowns, We need additional information. The directly measurable quantity is the radial velocity of the exocomet during transit. This can be determined by examining the spectra to identify the radial velocity corresponding to maximum absorption. We express the radial velocity as $\frac{dr}{dt}$. Referring to Eq.\,\ref{eq:parabolic}, we see that $r$ depends on $f$, an angle that varies with time and so we have that:

\begin{equation}
    \frac{dr}{dt} = \frac{dr}{df} \cdot \frac{df}{dt}
\label{eq:dr_dt}
\end{equation}

\noindent Starting with Eq.\,\ref{eq:parabolic} we differentiate (using the quotient rule for differentiation) to get:

\begin{equation}
    \frac{dr}{df} = \frac{2q\sin f}{(1+\cos f)^2}
\label{eq:dr_df}
\end{equation}

\noindent To calculate $\frac{df}{dt}$ we take advantage of the fact that the specific angular momentum $h$, which remains constant throughout the orbit, can for a parabolic orbit be expressed using Kepler's second law as:

\begin{equation}
    h = r^2 \frac{df}{dt} \Rightarrow \frac{df}{dt} = \frac{h}{r^2} 
\label{eq:df_dt}
\end{equation}

\noindent We note that $h$ can be expressed as

\begin{equation}
    h=\sqrt{2q\mu} 
\label{eq:h}
\end{equation}

\noindent where $\mu=GM_\star$. Eq.\ref{eq:h} shows that for a parabolic orbit, $h$ depends solely on the periastron distance $q$. Intuitively, this means that the farther an exocomet passes from the star, the greater its angular momentum is. Conversely, if the exocomet were on a purely radial path, its angular momentum would be zero. Substituting Eq.\,\ref{eq:dr_df} and Eq.\,\ref{eq:df_dt} into Eq.\,\ref{eq:dr_dt} we get:

\begin{equation}
    \frac{dr}{dt} = \frac{2q\sin f}{(1+\cos f)^2} \cdot \frac{h}{r^2} 
\label{eq:dr_dt_2}
\end{equation}

\noindent which after substituting in $r$ from Eq.\,\ref{eq:parabolic} and $h$ from Eq.\,\ref{eq:h} gives us after simplifying:

\begin{equation}
    \frac{dr}{dt} = \sqrt{\frac{\mu}{2q}}\sin{f} = \sqrt{\frac{GM_\star}{d}} \cdot \frac{\sin f}{\sqrt{1 + \cos f}} = - \sqrt{\frac{GM_\star}{d}} \cdot \frac{\sin \varpi}{\sqrt{1 + \cos \varpi}}
\label{eq:dr_dt_final}
\end{equation}

\noindent Keeping in mind that $\cos(f) = \cos(-\varpi) = \cos(\varpi)$. We find that the only term we don't have a value for is $\varpi$. We choose to express our results in terms of $\varpi$, as this angle defines the orbit and remains constant over time, unlike the true anomaly $f$. As a reminder $f=-\varpi$ only at transit. To calculate $\varpi$ requires solving the equation using a numerical inversion method as it cannot be solved analytically. Once a value for $\varpi$ is determined, we can use Eq.\,\ref{eq:parabolic} to calculate $q$, the periastron distance (keeping in mind that $\cos{f} = \cos{-\varpi} = \cos{\varpi}$). Currently, only exocomets in \bp\, have had their accelerations measured. This due to \bp\, being bright enough for high resolution measurements with sufficient signal-to-noise to measure accelerations. Additionally, \bp\, is unique for having a large number of transiting exocomets. It is important to note that there is a strong bias: only close-in exocomets have measurable accelerations, so the distances to more distant exocomets (from the star) cannot be determined due to the limitations in detecting accelerations in their spectra.

\subsection{Exocomets compared to Solar System comets}
Comets in the solar system exhibit a wide range of properties, both dynamically and compositionally, based on their formation locations within the protoplanetary disc. Understanding their origins (e.g., Kuiper Belt, Oort Cloud) and chemical compositions helps us establish the taxonomy of comets. This, in turn, provides insights into the composition and structure of the protoplanetary disc from which the solar system formed and into the dynamical interaction with planets at the early stage of the system.

Comparing exocomets directly to solar system comets poses many challenges. Solar system comets exhibit large variations in composition, with molecular abundances often varying by orders of magnitude (see Fig. 4 in \citealt{mumma_2011}). This variability likely extends to exocomets as well. The comparison is further complicated by how, when and where the comets are observed. The composition of solar system comets are measured for individual comets, either through direct observations or through in-situ measurements by satellites. The different parts of the comet (e.g. coma, tails and nucleus) can be studied individually. In contrast, exocomets are spatially unresolved. Their composition are inferred from various observations such as transit observations, analysis of polluted white dwarf atmospheres, or detection of gas produced through collisions within debris discs. Exocomets are never directly observed or visited. The behaviour of a comet depends on its environment. The proximity to a star (along with stellar characteristics) will determine what volatiles may be detected and how long they last before they are dissociated. A comet that has completed multiple orbits around its central star may be depleted in certain species, such as volatiles, compared to comets approaching the star for the first time. For a more in depth comparison, the reader is encouraged to look at \citep{strom_2020}.

\section{Exoasteroids}\label{exoasteroids}
The presence of exoasteroids, like exocomets, are inferred from indirect observations. A number of white dwarfs show photometric and spectroscopic signatures of planetesimals accreting onto their surfaces, which includes exocomets and exoasteroids.


Elements heavier than helium sink from the white dwarf (WD) envelope to its interior relatively fast, typically days to millions of years. When heavy elements are detected in the envelopes of white dwarfs with ages far exceeding their settling time, it suggests a more recent deposition, likely within the last few million years. Observing WDs provides us with the opportunity to measure the bulk composition of the accreting material. Fortunately, WDs showing recent accretion signatures are common. DA WDs, which are the most prevalent type of WDs in our galaxy (roughly 80\% of all white dwarfs), often display accretion signatures. A study by \cite{koester_2014} found that for DA WDs with cooling ages of 20-200 Myrs and 17 000 K $<$ T$_\mathrm{eff}$ $<$ 27 000 K) somewhere between a quarter and half show evidence for accretion planetary debris, which includes exoasteroids.

Metal lines are often detected in near-ultraviolet spectra obtained using the Hubble Space Telescope with the Space Telescope Imaging Spectrograph (STIS) and Cosmic Origins Spectrograph (COS) instruments. The accreted material is most often refractory-rich and volatile-poor, showing similarities with Solar system asteroids. The dominant species are the rock-forming elements O, Mg, Si and Fe with trace amounts of C and water \citep{zuckerman_2007,gansicke_2012}. There is also some evidence for more volatile-rich bodies being accreted, consistent with cometary material \citep{farihi_2013, xu_2017} although this is rare in comparison.

The source of the accreting material is thought to originate from circumstellar debris discs. They form once the exoasteroids survive the giant branch phase of the host star at distances of several au's before they are subsequently perturbed onto close-in orbits around the remaining WD, where they get disrupted. Once the exoasteroid gets within the Roche radius of the star it is tidally disrupted and accreted onto the WD. Such a tidal breakup was detected in WD\,0145+234 by the presence of a mid infrared outburst and was interpreted as a tidal disruption of an asteroid by \citep{wang_2019}. The white dwarf WD 1145+017 is a particularly interesting case as observations showed ongoing and variable accretion \citep{vanderburg_2015}.

\section{Interstellar visitors}

\subsection{\oum}
\begin{figure}[t]
\centering
\includegraphics[width=0.5\textwidth]{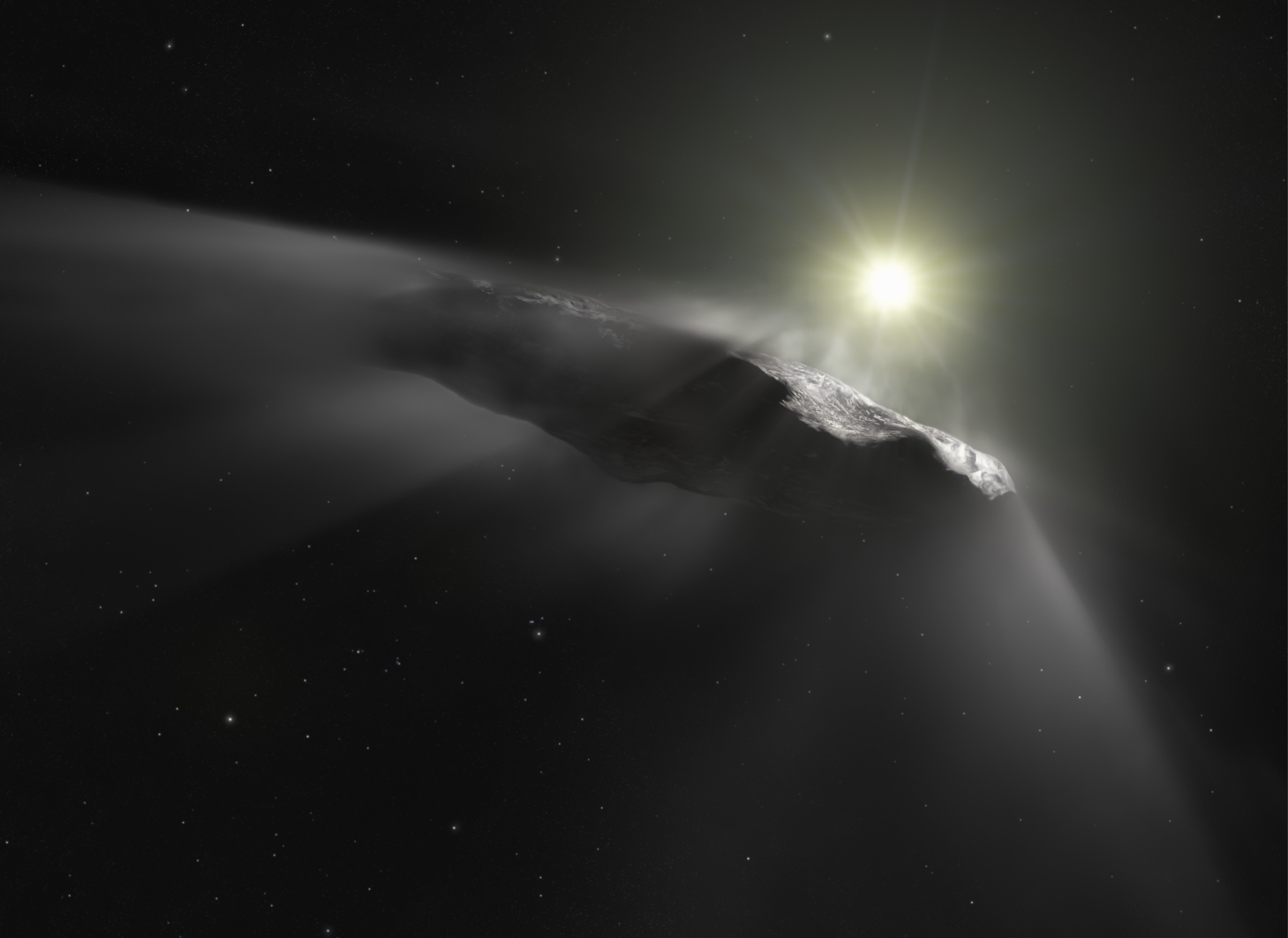}
\caption{An artists impression of \oum. Observations showed that the object is moving faster away from us than what was expected from gravitational acceleration. Image credit: ESA/Hubble, NASA, ESO, M. Kornmesser.}\label{img:oumuamua}
\end{figure}

The discovery of \oum\, by \cite{williams_2017} marked the first time a 100 m - kilometer scale small body originating from outside of our solar system was discovered traversing the inner solar system. During the early stages of planetary system formation, numerous small bodies known as 'planetesimals' are likely expelled from the systems where they formed. It was discovered by Robert Weryk as part of the Pan-STARRS project, which is a all-sky survey aimed at detecting small solar system bodies. Follow-up observations confirmed an eccentricity value of $e = 1.20113 \pm 2.1064 \times 10^{-5}$ which meant the orbit was hyperbolic. Scattering within the solar system was ruled out soon after and the interstellar origin was confirmed. It was first discovered after it had made its closest approach to the Sun (where it came within a distance of 0.255 au). Fig\,\ref{img:oumuamua} shows an artists impression of \oum.

From the outset, \oum\, did not show typical cometary characteristics. Instead, it showed a featureless red reflection spectrum with no detectable dust or gas tail or coma (despite using some of the worlds largest telescopes to look for them). Despite spectroscopic observations showing a featureless spectrum, they at least provided upper limits on the production rates \citep[e.g.][]{jewitt_2017, trilling_2018} of various volatile species and dust from \oum. This lack of activity made it compatible with an inert asteroid whose motion was driven primarily by gravity. However, it soon emerged that gravity alone could not explain its motion, and that something else was pushing it, creating a notable nongravitational acceleration similar to what is observed for gas and dust emitting comets.

\cite{bergner_2023} put forward the hypothesis that it is the release of entrapped molecular hydrogen which is responsible for the additional acceleration. They explain that \oum\, likely started off as a water-rich planetesimal resembling a comet. During its interstellar journey, it was irradiated by galactic cosmic rays and high-energy photons, which, through the process of radiolysis, split apart H$_2$O and produced entrapped H$_2$. Finally, the H$_2$ was released during the passage through the Solar System when subsurface layers warmed to temperatures where amorphous ice crystallizes. The more structured crystalline form results in the release of H$_2$ that was previously held within the amorphous ice matrix.

The nondetection of CO using Spitzer \citep{trilling_2018} ruled out a continuous acceleration driven by CO outgassing (although sporadic outgassing remains an option). There were no observations done which were sensitive to H$_2$0 features, however, as \cite{seligman_2020} point out, the nongravitational acceleration observed on its outbound trajectory is not consistent with water ice sublimation-driven jets, as observed for solar system comets. This is because the sunlight did not provide enough energy to power the acceleration of \oum\, via H$_2$0 sublimation (mainly due to the high latent heat of sublimation). \oum's observed properties can be explained if it contained a significant fraction of molecular hydrogen (H$_2$) ice. The incident solar flux produces a H$_2$ sublimation rate capable of producing the acceleration. This sublimation may also be what has lead to the elongated shape of \oum. Given that the ice processing leading to the outgassing of H$_2$ is most efficient in the top few meters of a comet, the acceleration will be more pronounced for smaller bodies, whereas for more massive bodies, the acceleration effect could be too small to be measured. Although the size of \oum\, is degenerate with its geometric albedo, it is clear that it is at least an order of magnitude smaller than Solar System comets (typically a few km in size), which display measurable activity indicators such as the presence of a tail and a coma. Estimates of the size of \oum\, vary from 55 to just over 100\,m. One way to put the theory of \cite{bergner_2023} to the test is to detect other small long-period and interstellar comets, then observe their orbits and look for nongravitational acceleration contributions. Ideally these observations would occur well before perihelia so that the effect could be observed before active sublimation of H$_2$O or CO/CO$_2$ occurs, as this effect could happen at much lower temperatures.

Photometric light curve observations of \oum\, showed extreme brightness variations varying by a factor $\sim12$, which went beyond what has previously been observed for most solar system objects. This, along with consistently observed deep minima in the light curve, suggests that 'Oumuamua has an oblate geometry, being flattened at the poles and bulging at the equator \citep{mashchenko_2019}. The faintness of \oum, in combination with it only being visible for a few weeks from most observatories, made it hard to constrain its dynamical and physical state. Photometric and spectral data in the optical to near-infrared showed no detection of any gas, including searches for CN, CO and CO$_2$. The red colour of \oum, which is consistent with an organic-rich surface and iron-rich minerals is thus not very constraining, showing that colour by itself is not a diagnostic of composition.

\subsection{2l/Borisov}

\begin{figure}[t]
\centering
\includegraphics[width=0.5\textwidth]{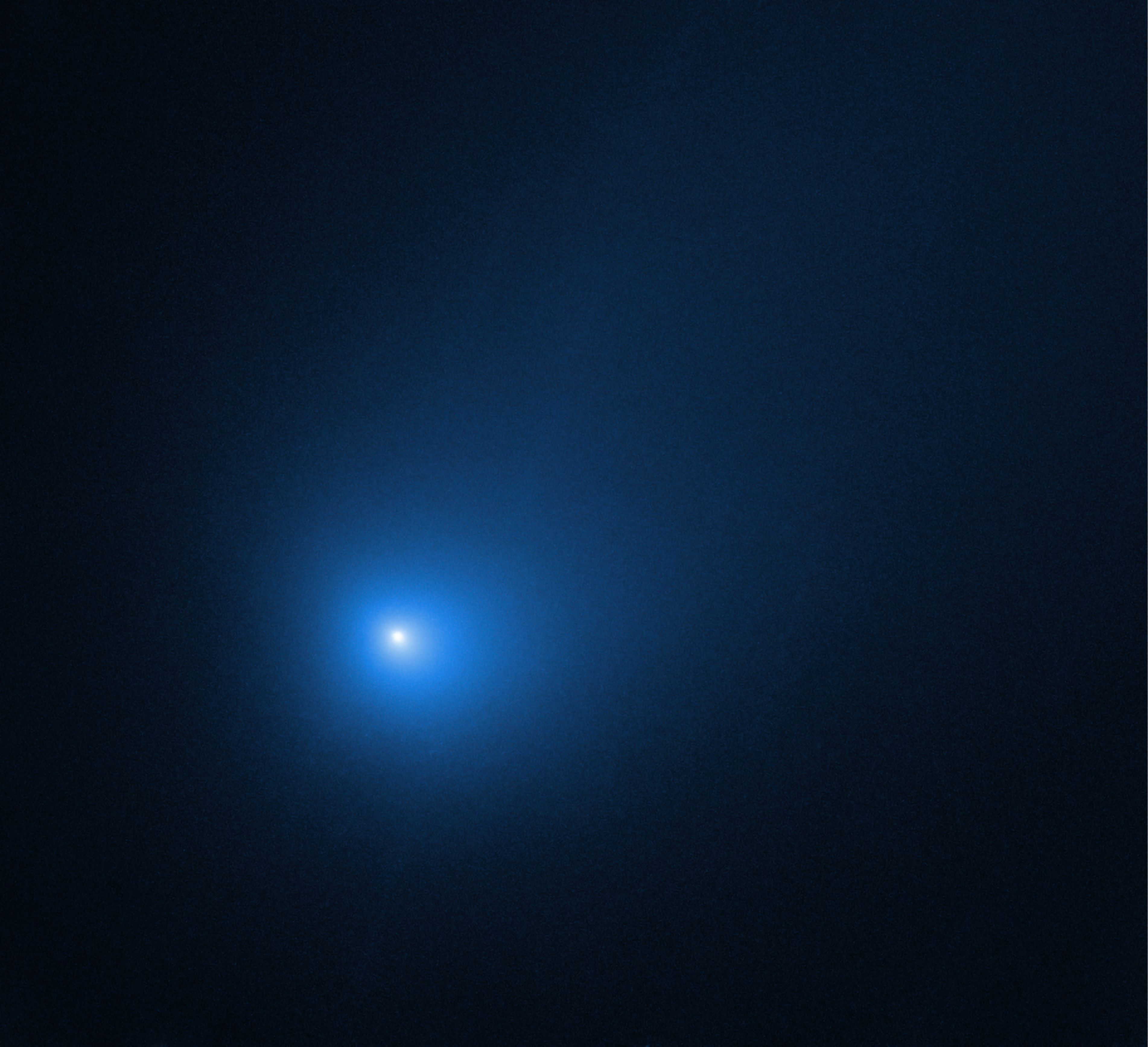}
\caption{An image of \2l\, taken with the NASA/ESA Hubble Space Telescope shortly after its closest approach to the Sun where it experienced intense heating after enduring the frigid conditions of interstellar space. Taken when the comet was at a distance comparable with that of the inner edge of the asteroid belt the comet itself is to small to be resolved in this image. The inner light blue part is the coma whereas the blue light around it comes from reflected dust. Comet \2l is only the second interstellar object known to have passed through our Solar System after \oum. Image credit: NASA, ESA, and D. Jewitt (UCLA).}\label{img:borisov}
\end{figure}

\2l, also known as C/2019 Q4 (Borisov), was discovered as an active comet by Gennadiy Borisov, amateur astronomer and engineer at the Crimean Astronomical Station. From the observed orbital properties it was clear that the object was of interstellar origin with an eccentricity of $e=3.3564758 \pm 1.9039 \times 10^{-5}$. As \2l was discovered $\sim3$ months before the perihelion passage ($q=2.0065 \pm 3.2904 \times 10^{-6}$\,au) left enough time for the mobilisation of observations. An example of these observations are those taken with the NASA/ESA Hubble Space Telescope (HST), which clearly depicts the cometary nature of \2l \citep{jewitt_2019, jewitt_2020}, see Fig.\,\ref{img:borisov}. Observations with the HST resulted in no direct detections of the nucleus, but did provide a strong upper limit on the radius of the nucleus $r\leq0.5$\,km. The nongravitational acceleration combined with an assumed density also provides a range of lower limits on nucleus radius. The shape of the comet remains largely unknown, in common with remote studies of solar system comets. The cometary coma has been studied in detail by several groups. High-resolution spectra at 12 different epochs from 2.1 au pre-perihelion to 2.6 au post-perihelion showed that \2l bears a remarkable compositional resemblance with solar system comets at optical wavelengths \citep{opitom_2021}. In particular the NiI to FeI abundance ratio was shown to be consistent with a sample of solar system comets such as Jupiter-family comets, Oort cloud comets, and Halley-type comets (HTC). Despite these similarities, when we look beyond the optical, targeting UV wavelengths, we see that \2l is not like most solar system comets. Observations of the coma with the HST and the COS spectrograph showed there was three times the amount of CO gas compared to H$_2$O gas when compared with comets in the inner ($<2.5$\,au) solar system \citep{bodewits_2020, cordiner_2020}. Comets which have CO as the dominant volatile, as opposed to H$_2$O, are very rare. Both Comet C/2016 R2 and 29P/Schwassman-Wachmann 1 have showed CO/H$_2\mathrm{O}\gg1$. However, in the case of 29P the high ratio can at least be fully explained by the comets' heliocentric distance being beyond the water ice line (where water is too far away from the Sun to sublimate) and so the composition of the coma may differ from that of the nucleus. The comparison is made with the inner solar system were both species can sublimate efficiently (further out CO$_2$ and CO may still sublimate efficiently, whilst H$_2$O does not).

Results from the Rosetta mission, which observed the 67/P-Churyumov-Gerasimenko solar system comet for two years, indicate the depth at which temperature variations penetrate into a material ranges from a few tens of cm to 1\,m. An interstellar comet is subjected to isotropic radiation primarily from cosmic rays and intermittent stellar sources. This radiation causes nonthermal desorption of surface volatiles through cosmic ray interactions and far-UV photon absorption, leading to the continuous erosion of its surface material. As a result, volatile materials are preferentially lost, suggesting that the CO/H$_2$O ratio in an interstellar comet may not accurately reflect the primordial ratio, even at depths of several tens of meters.

Using estimates that the comet lost its outer surface going down to a depth between 1.0 and 6.4\,m (depending on the size of the nucleus) suggests that as the comet passed perihelion, less altered layers of the comets may have been exposed. This would suggest the high CO/H$_2$O abundance ratio is either a primordial feature or that H$_2$O has been preferentially lost compared to CO and CO$_2$. For the feature to be primordial, the CO would have to have originated from a pristine portion of the object which since creation has stayed below the sublimation point of CO (typically around 25\,K in cometary nuclei). In this case the abundance of CO provides clues to the temperature of its formation environment and thermal history. There is no way of telling if the abundance ratio is primordial or not. Although it may be tempting to infer the plausible scenario that the protoplanetary disc where \2l formed was enriched in carbon relative to our own, the ratio may be the result of the preferential volatile loss or that the comet formed beyond the CO snowline within the system it formed.

\section{Exomoons}\label{chap1:sec7}%

To date, exomoons have proven elusive and difficult to detect despite their likely abundance, given that our own solar system contains nearly 300 moons. Although no bona fide detection has been made, several promising candidates exist, and significant advancements have been made in the last few decades in both understanding their occurrence around planets and refining the observational strategies used to search for them.  Several proposed exomoon detection methods exist, including detecting light curve variations caused by transits \citep{sartoretti_1999, kipping_2013}, microlensing \citep{bennett_2014}, polarimetry \citep{sengupta_2016}, transit timing variations \citep{kipping_2009} and direct imaging \citep{agol_2015, vanderburg_2018}. Within the transit methods, it is common to look for light curve shape alterations in the exoplanet light curve caused by the presence of an exomoon in orbit around an exoplanet which transits a host star. An exoplanet hosting an exomoon need not transit a star for it to be detected, however. Photometric monitoring of self-luminous, young, free floating planetary mass objects at near-IR wavelengths would allow exomoons to be detected either from variations in the light curve or through direct imaging observations \citep{limbach_2023}. Future space-based coronographs\footnote{Coronographs are instruments which block out the light from the star, allowing companion planets to be seen.} may be used to conduct direct imaging observations of exoplanets with companion moons that outshine the planet at certain wavelengths. The detection strategy involves measuring the position variation of the centre of light with wavelength, also known as \textit{spectroastrometry}. It has the potential to disentangle the spectral contribution from the exoplanet and the exomoon as well as determine the exomoon's orbit and the exoplanet mass.

Despite the many options available for detecting exomoons, there have so far only been a few candidates identified, especially when compared to the confirmed exoplanets, which number in their thousands. There are a number of reasons for this: the signals caused by exomoons are often very small and can easily be lost within the measurement uncertainties. They also do not show regular periodicities, given that the moon can show up before, during, or after the exoplanet transit. This makes it much harder to work with phase-folding and stacking techniques (both of which help increase the signal).

One of the early systematic searches was the HEK (Hunt for Exomoons with Kepler) project led by \cite{kipping_2012}. They surveyed the most promising moon-hosting candidates from the Kepler mission (in total 60 exoplanets). Then they compared the Bayesian evidence of a planet-only model to that of a planet-with-moon model with the goal of determining the frequency of large moons around viable planetary hosts, termed $\eta_{\text{\leftmoon}}$. To date, only two candidates emerge from the project as moon candidates: Kepler-1625 b-i and Kepler-1708 b-i\footnote{The \textit{b} indicates the planet (\textit{b} being given to the first planet detected around the star Kepler-1708) and \textit{i} the moon (moons are given Roman numerals with lower case numerals indicating candidate status).} Kepler-1625 b which had shown three exoplanet transits in the Kepler data was observed a fourth time with the HST allowing for a combined HST and Kepler data analysis of the four transits. \cite{teachey_2018}, who discovered the exomoon candidate, found that based on the presence of significant TTVs, and a sustained flux reduction in the HST light curve following planetary egress, the exomoon hypothesis remains the best explanation of the data. However, other groups called this detection into question, either through statistical arguments \citep{heller_2019}, or as an artifact of the data reduction \citep{kreidberg_2019}. \cite{kipping_2022} describes the detection of Kepler-1708 b-i, as a $\sim2.6$ Earth radii ($R_\oplus$) exomoon on an approximately coplanar orbit at $\sim12$ planetary radii from its Jupiter-sized host. At a distance of $\sim1.6$ au from the star, the period of the planet is very long at 737 days and as a result only has two transits present in the Kepler data. Future observations are necessary to validate the target, especially TTVs, which provide a second way of validating the target beyond studying the shape of the transit light curve. In both cases, the Kepler-1625 b-i and Kepler 1708 b-i exomoon candidates are large moons ($\sim0.5R_\oplus$ and $\sim2.6R_\oplus$ respectively) compared to Galilean-sized moons ($\sim0.2 – 0.4R_\oplus$). This could be due to large exomoons being more frequent, but more likely than not, it could be due to them being easier to detect (much in the same way that the first exoplanets detected were hot Jupiters, which we today know are relatively rare).
Microlensing surveys may very well have detected exomoons already. The MOA-2011-BLG-262 lensing event may have shown the presence of an exomoon \citep{bennett_2014}. Due to degeneracies in the model it remains uncertain if the lensing object consists of a distant red dwarf star with a planetary companion (18 times the mass of Earth) or a planet more massive than Jupiter which is closer to us with a sub-Earth-mass moon. Despite this unfortunate degeneracy, it was shown that microlensing measurements are indeed capable of detecting exomoons. 

\section{Conclusions}\label{sec:conclusions}%

In recent decades, the fields of exocomets, exoasteroids, interstellar visitors, and exomoons have seen remarkable growth. The future looks bright with new telescopes holding the potential for many new discoveries to come. The James Webb Space Telescope (JWST) will enable detailed characterisation of exocometary material at longer wavelengths, where molecular signatures are most prominent. PLATO, ESA's upcoming exoplanet mission, offers high photometric precision and a wide field of view, which will significantly expand the sample of known exocomet host stars and enhance our ability to detect exomoons. Additionally, the Vera Rubin Observatory is set to revolutionise the detection of transient phenomena, greatly increasing our ability to identify interstellar visitors. It's an exciting time to be involved in these rapidly advancing fields, as new technologies promise to deepen our understanding of the universe.

\begin{ack}[Acknowledgements]

I would like to thank Grant Kennedy, Darryl Seligman, Alain Lecavelier des Etangs, Tim D. Pearce, Th{\'e}o Vrignaud and Dimitri Veras for their valuable input. Fig\,\ref{fig:orbit} is adapted based on work by Lasunncty and sourced from Wikipedia, licensed under the Creative Commons Attribution-Share Alike 3.0 Unported license.

\end{ack}


\bibliographystyle{Harvard}
\bibliography{reference}

\end{document}